\documentclass[aps,prb,twocolumn,nopacs,amsmath,amssymb]{revtex4-1}
\usepackage{bm}
\usepackage{times}
\usepackage{subfigure}
\usepackage{graphicx}
\usepackage{multirow}
\usepackage{hyperref}
\usepackage[dvipsnames]{xcolor}
\begin{document}
\title{Domain walls in a chiral $d$-wave superconductor on the honeycomb lattice}
\author{Oladunjoye A. Awoga}
\author{Adrien Bouhon}
\author{Annica M. Black-Schaffer}
 \affiliation{Department of Physics and Astronomy, Uppsala University, Box 516, S-751 20 Uppsala, Sweden}
% \date{\today}
\begin{abstract}
We perform a fully self-consistent study of domain walls between different chiral domains in chiral $d_{x^2-y^2} \pm id_{xy}$-wave superconductors with an underlying honeycomb lattice structure. 
We investigate domain walls along all possible armchair and zigzag directions and with a finite global phase shift across the domain wall, in addition to the change of chirality.
For armchair domain walls we find the lowest domain wall energy at zero global phase shift, while the most favorable zigzag domain wall has a finite global phase shift dependent on the doping level. Below the van Hove singularity the armchair domain wall is most favorable, while at even higher doping the zigzag domain wall has the lowest energy. The domain wall causes a local suppression of the superconducting order parameter, with the superconducting recovery length following a universal curve for all domain walls.
Moreover, we always find four subgap states crossing zero energy and well localized to the domain wall. However, the details of their energy spectrum vary notably, especially with the global phase shift across the domain wall.
\end{abstract}
\maketitle
% ------------------------------------------- %
% INTRODUCTION:
% ------------------------------------------- %
\section{Introduction}
% Introduce topo SC, chiral SC:
Superconductors with a non-trivial chiral topology have, and continue to, generate a lot of interest.\cite{VolovikGorkov,mackenzieandManoe2003, HassanKane2010RMP,qi2011topological, kallin2012chiral,BlackSchafferandHonerkamp2014,kallin2016chiral} 
Early prominent examples includes the proposed bulk spin-triplet $p_x\pm ip_y$-wave state in Sr$_2$RuO$_4$\cite{mackenzieandManoe2003, kallin2012chiral,kallin2016chiral} and a spin-singlet $d_{x^2-y^2}\pm id_{xy}$-wave state close to the surface or at impurities in the high-temperature cuprate superconductors, \cite{Volovik97,fogelstrom1997tunneling, Covington97, Krishana97,Laughlin98, balatsky1998spontaneousTRB} and more recently also in heavily doped graphene.\cite{Black-Schaffer07,nandkishore2012chiral,Wang11, Kiesel12, BlackSchafferandHonerkamp2014}  There also exist proposals for chiral superconductivity in heavy fermion superconductors, \cite{VolovikGorkov} with recent experiments finding evidence for a chiral spin-triplet $f_{(x^2-y^2)z} \pm 2 i f_{xyz}$ in UPt$_3$ in the low-temperature B phase.\cite{joynt2002superconducting,strand2009evidence,strand2010transition,schemm2014observation}

% DWs:
In a chiral superconductor there is no energy difference between the two different chiralities set by the sign between the two order parameter components. These are simply the two different but equivalent choices for the superconducting order parameter when forming the intrinsically complex order parameter.
This energy degeneracy between different chiral states naturally allows for domain wall formation inside the superconductor, between domains with different chiralities. Domain walls thus play an important role for the properties of chiral superconductors. In fact, for the $p_x+ip_y$-wave state in Sr$_2$RuO$_4$ domain walls have already been rather extensively investigated.\cite{matsumoto1999quasiparticle,kidwingira2006dynamical,bouhon2010influence} For example, domain wall formation naturally limits the measurable edge currents carried by the edge states generated by the chiral topology. \cite{kirtley2007upper,hicks2010limits,curran2014search} 

% Spin-singlet chiral SCs:
Chiral states have also gained increasing attention in spin-singlet superconductors. Notably, there exists vastly more known spin-singlet superconductors, ranging from conventional electron-phonon $s$-wave superconductors to strongly-correlated superconductors with $d_{x^2-y^2}$-wave symmetry with the cuprates as the most prominent example.\cite{Tsuei00RMP} In fact, at surfaces of cuprate superconductors a subdominant $i d_{xy}$ component has been proposed to be present in order to explain experimental results, such as a fully gapped state in cuprate nano-islands,\cite{Gustafsson13, Black-Schaffer13} split zero-energy conductance peaks in tunneling experiments,\cite{Covington97, fogelstrom1997tunneling} and thermal conductivity measurements.\cite{Krishana97}
Even more generically, for $d$-wave superconductors with an underlying three- or six-fold lattice symmetry, the two different $d$-wave symmetries, $d_{x^2-y^2}$ and $d_{xy}$, are necessarily degenerate. This very generically leads to a $d_{x^2-y^2}+id_{xy}$-wave chiral superconducting order parameter in such materials.\cite{Black-Schaffer07,nandkishore2012chiral,BlackSchafferandHonerkamp2014}

% Examples of honeycomb/triangular SCs:
Multiple different materials have already been proposed to be intrinsic $d_{x^2-y^2}+id_{xy}$ superconductors due to their six-fold symmetric honeycomb lattice structure.\cite{BlackSchafferandHonerkamp2014} Graphene, heavily doped to the van Hove singularity (VHS), is one example, where multiple (functional) renormalization group (f)RG calculations have found a $d_{x^2-y^2}+id_{xy}$-wave state due to electron-driven superconductivity.\cite{nandkishore2012chiral, Wang11, Kiesel12}
The superconducting pnictide SrAsPt with $T_c = 2.4$~K \cite{Nishikubo11} is also composed of honeycomb layers. Here recent muon spin-rotation ($\mu$SR) experiments\cite{biswas13} have revealed time-reversal symmetry breaking, which is in agreement with fRG calculations finding a chiral $d$-wave state.\cite{Fischer14}
Also In$_3$Cu$_2$VO$_9$ \cite{moller2008structural, yan2012magnetic, liu2013interlayer, Wu13, Black-Schaffer&LeHur14} and several different irridates,\cite{okamoto2013dopedPRL, okamoto2013globalPRB} all with an effective honeycomb structure, have been proposed to host a $d_{x^2-y^2}+id_{xy}$-wave superconducting states upon doping the magnetic ground state.  

% Consequences of d+id:
Despite the growing interest and evidence for spin-singlet $d_{x^2-y^2}+id_{xy}$-wave superconductivity, relatively little is known about the properties of this state. The $d_{x^2-y^2} \pm id_{xy}$ state has a Chern\cite{BlackSchafferandHonerkamp2014, tanaka2011symmetry}, or winding, number $C = \pm 2$, which automatically gives rise to two topologically protected and chiral states at subgap energies at each edge.\cite{Volovik97, BlackSchafferandHonerkamp2014}
Recently, a more careful study of the edge states of a $d_{x^2-y^2}+id_{xy}$-wave superconductor on the honeycomb lattice has shown that the $d_{xy}$ component is actually heavily suppressed on all types of edges due to pair breaking effects.\cite{Black-Schaffer12PRL} Despite this pair breaking, edge states were still found and also well localized to the edges. 
For a domain wall between the two different chiralities of the chiral $d$-wave state, the bulk-edge correspondence\cite{Volovik97, HassanKane2010RMP, qi2011topological, graf2013bulk} likewise dictates a total of four subgap states at the domain wall, since the Chern number changes by four across the wall.\cite{Volovik97, BlackSchafferandHonerkamp2014} Very recent work has also calculated the total domain wall subgap spectrum semiclassically, confirming the existence of domain wall states.\cite{mukherjee2016quasiparticles}
However, beyond this, very little is known about either the explicit domain wall structure or the detailed properties of the subgap domain wall states. 
For example, the same mismatch between the honeycomb lattice and the $d$-wave symmetry giving rise to the chiral $d$-wave state in the first place, also allows for multiple possible crystal orientations for domain walls in honeycomb superconductors, which might have very different properties.

% What we do:
In this work we provide a fully self-consistent study of the superconducting state at generic domain walls in chiral $d$-wave superconductors with a honeycomb lattice structure.
More specifically, we study domain walls formed along all different zigzag (ZZ) and armchair (AC) directions of the honeycomb lattice and also allow for a finite global phase shift across the domain wall, in addition to the change of chirality. We find that the domain wall energy density depends not only on edge direction but also strongly on the global phase shift and the doping level.
For domain walls along the AC direction, the lowest energy is found for $x$-direction domain walls, with no additional global phase shift. This is also the overall most stable domain wall configuration for doping levels below the VHS.
For the ZZ direction, the lowest domain wall energy instead occurs at a finite phase shift that changes with doping. Beyond the VHS, where ZZ domain walls have the overall lowest energy, the phase shift is around, but not exactly equal to, $2\pi/3$.
We also find a finite suppression of superconductivity in the vicinity of the domain wall. We show that domain walls with a narrower suppression profile are generally more energetically favorable. But notably, we find that non-self-consistent results with a sharp order parameter profile at the domain wall often predict the wrong lowest energy domain wall configurations. 
We also establish that the recovery length for superconductivity away from the domain wall is inversely related to the order parameter strength in the bulk and not dependent on domain wall orientation.
We also study the induced subgap domain wall states. There are always four subgap domain wall states but their dispersion vary significantly with domain wall orientation and global phase shift. 
These results are directly applicable to the $d$-wave superconducting state in heavily doped graphene, but also to other chiral $d$-wave honeycomb superconductors.
%
% Organization of article:
The rest of the article is organized as follows: In Section II we introduce our model and discuss the bulk $d$-wave state.  In Section IIIA we analyze all possible domain walls and establish the most  energetically favorable configurations. Then in Section IIIB we determine the properties of the most favorable domain walls, focusing both on the domain wall subgap states and the order parameter at the domain wall. In Section IIIC we establish the generality of the superconducting recovery length. Finally, in Section IV we conclude our work.
% ------------------------------------------- %
% METHOD:
% ------------------------------------------- %
\section{Method}
%Figure of the lattice
\begin{figure}[ht!]	
	\centering	
  	\includegraphics[]{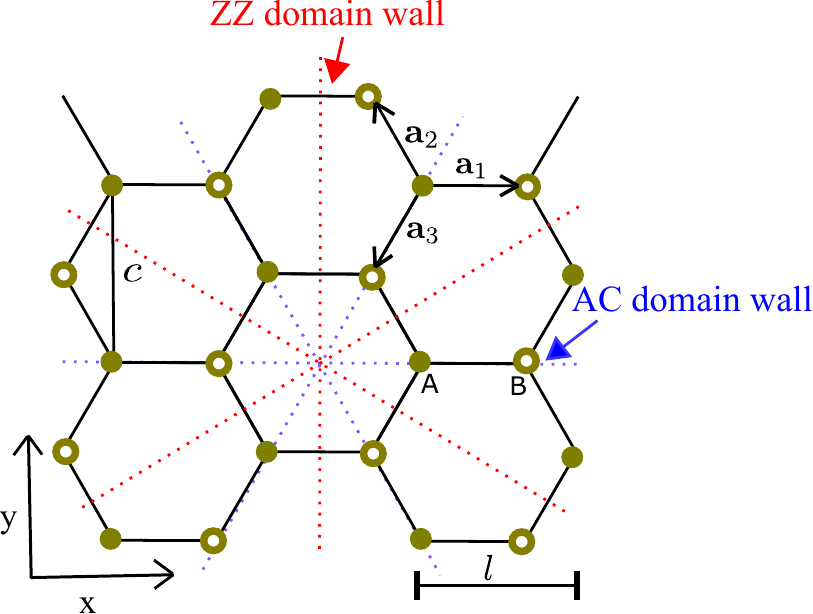}
	\caption[Honeycomb lattice]{(Color online.) Schematic diagram of the honeycomb lattice showing the atomic sites in sublattice A (filled circle) and sublattice B (circle) as well as the three nearest neighbor bonds directions $\mathbf{a_1}$, $\mathbf{a_2}$, $\mathbf{a_3}$. $c$ is the lattice constant and $l$ the distance between sublattice sites in the horizontal direction. Dotted lines show all three ZZ (red) and AC (blue) domain wall directions.
	}
	\label{fig:Lattice}
\end{figure}
%
% Model:
Many of the materials predicted to be chiral $d$-wave superconductors have a honeycomb lattice structure.\cite{BlackSchafferandHonerkamp2014} In order to keep the model as simple as possible, while still addressing general phenomena, we consider the honeycomb lattice with one ($s$-wave) orbital per site, including up to nearest neighbor processes. The general Hamiltonian for this system is $\hat{H}= \hat{H_0} + \hat{H}_\Delta$ with
%
%Hamiltonian
\begin{align}\label{eq:Hamiltmf} 
\hat{H}_0&=-t\sum_{i,p,\sigma}(a_{i\sigma}^\dag b_{i+\mathbf{a}_p\sigma}+\textrm{h.c.})
    	+ \mu\sum_{i \sigma}(a_{i\sigma}^\dag a_{i\sigma}+b_{i\sigma}^\dag b_{i\sigma}) \nonumber \\                                                                                                                                                                                                                                                                                                                                                                                                                                                                                                          
\hat{H}_\Delta&= \sum_{i,p} \lbrace\Delta_p(i) (a_{i\uparrow}^\dag    
             b_{i+\textbf{a}_p\downarrow}^\dag-a_{i\downarrow}^\dag b_{i+\textbf{a}_p\uparrow}^\dag)  
              + \textrm{H.c.}\rbrace 
    \nonumber
    \\
    &+\sum_{i,p}\frac{|\Delta_p(i) |^2}{J}.
\end{align}
Here $i$ is the site index and $\textbf{a}_p$ with $p = 1,2,3$ are the nearest neighbor bond directions, while $a_{{i}\sigma}^\dag (b_{{i}\sigma}^\dag)$ creates a particle with spin $\sigma$ at site $i$ in sublattice A (B). The lattice is shown schematically in Fig.~\ref{fig:Lattice}. 
$\hat{H}_0$ gives the normal state electronic band structure of the honeycomb lattice, essentially doped graphene, with hopping strength $t$ and chemical potential $\mu$. Since there is a Dirac point at $\mu =0$, we assume a finite doping in order to achieve superconductivity. The superconducting transition temperature will generally increase with doping as more density of states (DOS) appear at the Fermi level. However, at the VHS at $\mu=t$ the spectrum changes from two Fermi surfaces centered around $K,K'$ to one surface around $\Gamma$, which can influence many properties of the system. We therefore investigate the behavior of domain walls both below and above the VHS.

% SC:
Considering the highly unconventional nature of the chiral $d$-wave state, we use short-range superconducting pairing as would be generated by electron-electron interactions and also found to be present in e.g.~superconducting graphene close to the VHS.\cite{nandkishore2012chiral,Black-Schaffer07} We here limit the study to nearest-neighbor pairing since next-nearest-neighbor terms have been shown to cause only very limited modifications to the physics both at edges and around impurities.\cite{Black-Schaffer12PRL, Lothman14}
Within a fully self-consistent mean-field treatment the superconducting order parameter $\Delta_{p}(i)$ is given by
%
%Self-consistent equation
\begin{equation}\label{SelfConsist}
	\Delta_p(i)=\sqrt{2}J\langle a_{i\downarrow} b_{i+\mathbf{a}_p\uparrow} -a_{i\uparrow} b_{i+\mathbf{a}_p\downarrow} \rangle,
\end{equation}
where $J$ is the effective pairing strength. While $J$ is constant, the strong inhomogeneity produced by domain walls will generally render $\Delta_p(i)$ dependent on both lattice site and bond directions, and a fully self-consistent treatment of superconductivity is therefore necessary. Collating the three components of the order parameter as a vector, we write the order parameter as $\boldsymbol{\Delta} = (\Delta_1 \ \Delta_2 \ \Delta_3)$.

%BULK:
In the bulk the Hamiltonian Eq.~\eqref{eq:Hamiltmf} has two different superconducting solutions with different $T_c$ depending on $\mu$ and $J$.\cite{Lothman14}  One solution is an extended $s$-wave state, $s_\text{ext}$, which is favorable only at very high $J$ or $\mu$, but otherwise subdominant. The second solution consists of two degenerate $d$-waves, $d_{x^2-y^2}$ and $d_{xy}$, which appears at finite doping for a finite $J$. These solutions are spanned by the $D_{6h}$ irreducible representation basis vectors
%
% Basis vectors for D6h irrep for s and d- waves
\begin{align}\label{eq:bulksol}
	\hat{\boldsymbol{\Delta}}_{s_\text{ext}}& = \frac{1}{\sqrt{3}}\left(
	\begin{array}{ccc}
		1 & 1 & 1
	\end{array}
	\right), \nonumber  \\
	\hat{\boldsymbol{\Delta}}_{d_{x^2-y^2}} &=  \sqrt{\frac{2}{3}}\left(
	\begin{array}{ccc}
		1 & -\frac{1}{2} & -\frac{1}{2}
	\end{array}
	\right), \nonumber   \\
	\hat{\boldsymbol{\Delta}}_{d_{xy}} & =  \frac{1}{\sqrt{2}}\left(
	\begin{array}{ccc}
		0 & 1 & -1
	\end{array}
	\right).
\end{align}
$s_\text{ext}$ belongs to the one-dimensional $A_{1g}$ irreducible representation of the $D_{6h}$ point group, while the $d$-wave states belong to the two-dimensional doubly degenerate $E_{2g}$ irreducible representation.\cite{Black-Schaffer07} Very generally, such degeneracy dictated by group theory gives rise to a complex combination of the two symmetries having the lowest energy at temperatures below $T_c$. Indeed, it has been shown that below $T_c$ and at finite doping the honeycomb lattice favors either of two chiral $d$-wave superconducting states,\cite{Black-Schaffer07,BlackSchafferandHonerkamp2014, nandkishore2012chiral, Wang11, Kiesel12} which is an equal complex mixture of the two $d$-waves written as $d_{x^2-y^2} \pm id_{xy}$ with basis vectors
%
% Chiral d-wave vector
\begin{equation}\label{eq:chiral}
	\hat{\boldsymbol{\Delta}}_{d_{x^2-y^2}\pm id_{xy}}= \frac{1}{\sqrt{3}}\left(
	\begin{array}{ccc}
		1 &  e^{\pm i\frac{2\pi}{3}}  & e^{\mp i\frac{2\pi}{3}}
	\end{array}
	\right). 
\end{equation}
These chiral $d$-wave states are energy degenerate, time-reversal symmetry breaking, and fully gapped superconducting states. Calculating the Chern, or winding, number results in $C=\pm 2$.\cite{Volovik97, tanaka2011symmetry, BlackSchafferandHonerkamp2014}  This finite Chern number can easily be understood by noting that a $d_{x^2-y^2} + i d_{xy}$-wave order parameter winds $4\pi$ around the Brillouin zone center and thus has a winding of 2. At low doping levels, where the Fermi surface is instead centered around $K$ and $K'$, the winding can alternatively be considered on each separate Fermi surface where it is $2\pi$, thus adding up to again a total rotation of $4\pi$. Moreover, since the chiral $d$-wave state is fully gapped even when the Fermi surface topology changes at the VHS at $\mu = t$, the Chern number has to be constant on both sides of the VHS.
The sign of the winding number (or sign of the imaginary part of the order parameter) is referred to as the chirality or handedness of the order parameter. The energy degeneracy between the $C = \pm 2$ states easily results in domain formation inside the material. At the domain wall separating two different chiral domains the order parameter is expected to vary strongly. In order to capture this spatial variation we calculate $\boldsymbol{\Delta}$ self-consistently and we also define the $\Delta$-character of the local symmetry as $\boldsymbol{\Delta}(i)\cdot\hat{\boldsymbol{\Delta}}_r/|\boldsymbol{\Delta}(i)|$, where $\hat{\boldsymbol{\Delta}}_r$ represents the basis vectors in Eq.~\eqref{eq:bulksol}.

%------------------------------------%
% DOMAIN WALL
%------------------------------------%
\subsection{Domain wall configurations}\label{sec:DW}
The honeycomb lattice has two high-symmetry directions, the ZZ and AC directions, see Fig.~\ref{fig:Lattice}. We will consider domain walls along both of these directions. The most general form of the order parameter across a domain wall have both different chiralities and different global phases on opposite sides of the domain wall. For a sharp domain wall, that is a fixed order parameter domain wall, this can be written in the form
%
%Sharp DW order parameter
	\begin{align}\label{eq:DW}
	\boldsymbol{\Delta}	(i) &= \begin{cases}
			e^{i\phi_L}\boldsymbol{\Delta}  \qquad 0 < i \leq \frac{N}{2} \\
				e^{-i\phi_R}\boldsymbol{\Delta}^* \qquad \frac{N}{2} < i \leq N,	
						    \end{cases}  
	\end{align}
where $\boldsymbol{\Delta^{(*)}}$ has $d_{x^2-y^2} + id_{xy} (d_{x^2-y^2} - id_{xy})$ symmetry. Here $N$ is the number of unit cells in the direction perpendicular to the domain wall and $\phi_{L/R}$ is the global phase on the left/right side of the domain wall. We will use the parameter $\alpha = \phi_L+\phi_R$ to characterize the global phase shift across the domain wall. Thus, when $\alpha = 0$ there is a $\pi$-phase shift in $d_{xy}$ across the domain wall, while for $\alpha = \pi$ there is a $\pi$-phase shift in $d_{x^2-y^2}$ (and as it turns out also in $s_\textrm{ext}$) across the domain wall. 
Note that the domain wall form in Eq.~\eqref{eq:DW} naturally includes all three ZZ (AC) directions, shown in Fig.~\ref{fig:Lattice}, despite us only using the $\boldsymbol{\Delta}$ form for domain walls along by the $y$- $(x)$-axis. This is achieved by changing $\alpha$ by $\tfrac{2\pi}{3}$, which gives the required cyclic permutation of the bond order parameters for the remaining directions.

Due to the different geometry of the ZZ and AC directions, we need to define a domain wall energy density $\varepsilon$ which is the domain wall formation energy normalized by domain wall length. The formation energy is calculated as the difference between the free energy of the domain wall state and the free energy of the pure chiral state. $\varepsilon$ is thus a system size independent quantity which is needed in order to compare domain wall behavior along different lattice directions. 
Similarly, due to the difference in  distance between sites of the same sublattice along different directions we use normalized lengths, $\tilde{x}=x/l$ and $\tilde{y}=\sqrt{3}y/l$ (where $l$ is the horizontal distance between the same atomic species, see Fig.~\ref{fig:Lattice}) along $x$ and $y$-axes, respectively.

%------------------------------------%
%NUMERICALS
%------------------------------------%
\subsection{Numerical details}
We assume straight domain walls, see Fig.~\ref{fig:Lattice}, such that the system is translationally invariant in the direction parallel to the domain wall. This assumption allows for the use of periodic boundary condition where we Fourier transform into $k$-space in the direction parallel to the domain wall, which significantly reduces the size of the problem. We start the self-consistent calculations with an initial constant value of $\boldsymbol{\Delta}$ in pure chiral forms on both sides of the domain wall and use a gauge where $\phi_L = \phi_R$ such that only $\alpha$ is relevant. Diagonalizing Eq.~\eqref{eq:Hamiltmf} using this initial guess, we recalculate the order parameter using Eq.~\eqref{SelfConsist} at zero temperature, see e.g.~Ref.~[\onlinecite{Black-Schaffer2008self}]. We repeat these steps until the absolute difference in order parameters between two subsequence iterations is less than a small predetermined values on all sites (here set to $10^{-5}t$), which then signals self-consistency. To find the most energetically favorable configuration we do not need to impose any constraints during the self-consistency procedure. However, when we are comparing different $\alpha$ values which are not necessarily the ground state, we need fix the phase of $\boldsymbol{\Delta}$ on five lattice sites in each domain far from the domain wall and outer edges, in order to not change $\alpha$ during the self-consistency procedure.
In terms of system size $N$ we have ensured that bulk conditions are reached far from the domain wall and outer edges. This usually results in $N$ between 140 and 400, depending on the strength of the superconducting state. We have also checked for convergence with respect to the number of $k$-points, using normally up to 351 points, where an odd number of $k$-points ensures that the $\Gamma$-point of the Brillouin zone is also sampled.
In what follows, energies and lengths are, respectively, given in units of $t$ and $c$ (where $c$ is the lattice constant, see Fig.~\ref{fig:Lattice}), such that $\varepsilon$ is in units of $t/c$. 

% ------------------------------------------- %
% RESULTS:
% ------------------------------------------- %
\section{Results}
\subsection{Domain walls in general directions}
\begin{figure}[htb]	
	\centering	
	\includegraphics[width =8.6 cm]{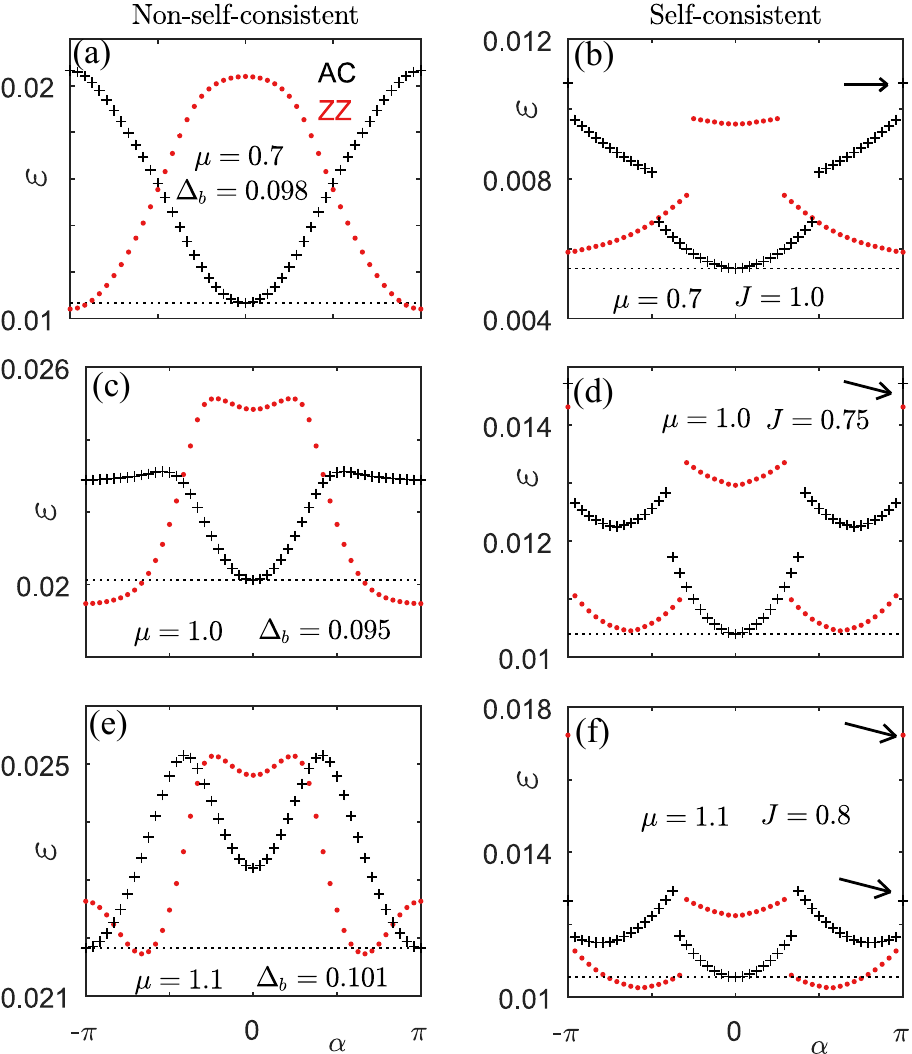}
	\caption[SFree energy]{(Color online.) Domain wall energy density $\varepsilon$ as a function of $\alpha$ for several $\mu$ and $J$ for sharp (a,c,e) and self-consistent (b,d,f) domain walls. Dotted horizontal line indicates the global minimum for the AC domain wall. For the sharp domain walls the bulk value of $\Delta_b = |\boldsymbol{\Delta}|$ are the same as in the self-consistent calculations and given in each subfigure. Arrows indicate hard to see data points at $\alpha=\pi$.}
	\label{fig:FvsAlpha}
\end{figure}

We first focus on how the domain wall energy density $\varepsilon$ varies both with the system parameters, $J$ and $\mu$, and different domain wall configurations, both ZZ and AC directions and for varying $\alpha$. Note that $J$ only sets the strength of superconductivity and should not directly play any major role. However, the Fermi surface of the system changes with $\mu$. Our numerical results shows that the overall behavior of $\varepsilon$ is indeed not affected by $J$ (see below in Fig.~\ref{fig:FvsMU}), but changes quite drastically as a function of domain wall direction (AC or ZZ), $\alpha$, and $\mu$, as seen in Fig.~\ref{fig:FvsAlpha}. 
Although $\varepsilon$ is both symmetric and periodic in $\alpha$, $\varepsilon(\alpha) = \varepsilon(-\alpha)= \varepsilon(\alpha+2\pi)$, the number of stable domain wall solutions and the global energy minimum $\varepsilon_\text{min} = \varepsilon(\alpha_ \text{min})$ change for different parameters and configurations. 
%
%Figure of free energy as a function of phase shift
%

%-------------------------------------------------------------%
% non-self-consistent solution of domain wall energy
%-------------------------------------------------------------%
Let us first briefly comment on the properties of sharp, and thus non-self-consistent, domain walls. Fig.~\ref{fig:FvsAlpha}(a,c,e) shows that domain walls along the ZZ direction (red dot) is always favored, but often not significantly, over the AC (black cross) direction. Another important feature is the change in $\alpha_\textrm{min}$ as the doping $\mu$ changes the topology of the normal state Fermi surface. For both the AC and ZZ domain walls there is a large shift in $\alpha_\textrm{min}$ at the VHS at $\mu = 1$, when the Fermi surface goes from two pockets at $K$ and $K'$ to only one at $\Gamma$. Note however that the chiral $d$-wave state is fully gapped in the bulk for all doping levels $-3t<\mu<3t$ and thus the Chern number and the number of domain wall states do not change with doping. Still, the domain wall structure is very sensitive to the normal state Fermi surface.

%-------------------------------------------------------------%
%self consistent solution of domain wall energy
%-------------------------------------------------------------%
Since sharp domain wall configurations are only a crude approximation, we turn to our main focus of self-consistent results. We find that self-consistent calculations significantly modify the behavior of $\varepsilon$. First, self-consistency lowers $\varepsilon$, as seen in Fig.~\ref{fig:FvsAlpha}(b,d,f), compared to a sharp domain wall. This is a natural result as the system is now allowed to relax to find a lower energy minimum. However, this result is still important as it shows that the system does not favor a sharp domain wall order parameter profile.
An even more interesting outcome of self-consistency is that the extrema of $\varepsilon$ are notably changed compared to the sharp domain scenario. This shows that self-consistency is very important to accurately capture not only finer details, but the overall configurations of domain walls in chiral $d$-wave superconductors. We can classify the different stable domain wall configurations into three groups. There are domain wall stable configurations at $\alpha=0$, called DWI, another group DWII exists at $\tfrac{\pi}{4} < \alpha < \tfrac{3\pi}{4}$, and the third group, DWIII, appears at $\alpha=\pi$. For AC domain walls, the DWI configuration is always most favorable in the self-consistent results (although notably not for sharp domain walls), although DWII solutions are also meta-stable at doping levels at or close to the VHS. This means that the AC domain walls prefer to be along the $x$-direction and with the order parameter symmetry $d_{x^2-y^2}\pm id_{xy}$.
On the contrary, for the ZZ direction DWI is always only meta-stable, while the global energy minimum is found for DWIII at low doping levels and for DWII for doping at and beyond the VHS. These latter solutions have a non-trivial rotation between the order parameter symmetry and the ZZ direction.
We also note that for the self-consistent solutions, there are clear discontinuities in the free energy between the different sets of stable configurations. These represent extremal and unstable maximum energy solutions.
The self-consistency results show a rather close energy competition between the AC and ZZ domain wall directions. At low doping levels domain walls along the AC direction are slightly favored within the DWI configuration, while ZZ domain walls with DWII configuration is preferred beyond the VHS. In the vicinity of the VHS both directions have the same energy within numerical accuracy. Again, these results are different from the sharp interface results where ZZ domain walls are always preferred. 

The solutions within the DWII group are particularly interesting. Although here $\alpha_\textrm{min}$ is close to the high-symmetry value $\alpha = 2\pi/3$ (only indicating an overall permutation of the order parameters between the three bonds), it varies with $\mu$. This shows that the preferential domain wall configuration very often appears with a finite $\alpha$. Only at the VHS for the ZZ domain wall is $\alpha_\textrm{min}= \tfrac{2\pi}{3}$.
These DWII solutions also have a built-in degeneracy between $\pm \alpha$. Such multiple non-universal minima for the domain wall energy, dependent on the interface orientation, have earlier also been predicted in chiral $p$-wave superconductors on the square lattice.\cite{sigrist1999role} In that case the appearance of two minima was also found to give rise to fractional vortices connected with the domain walls. Here we do not pursue the study of vortices at the domain wall of chiral $d$-wave superconductors, but leave this intriguing possibility to future studies. 

%------------------------------------%
%Strength of superconductivity
%------------------------------------%
\begin{figure}[htb]	
	\centering	
	\includegraphics[]{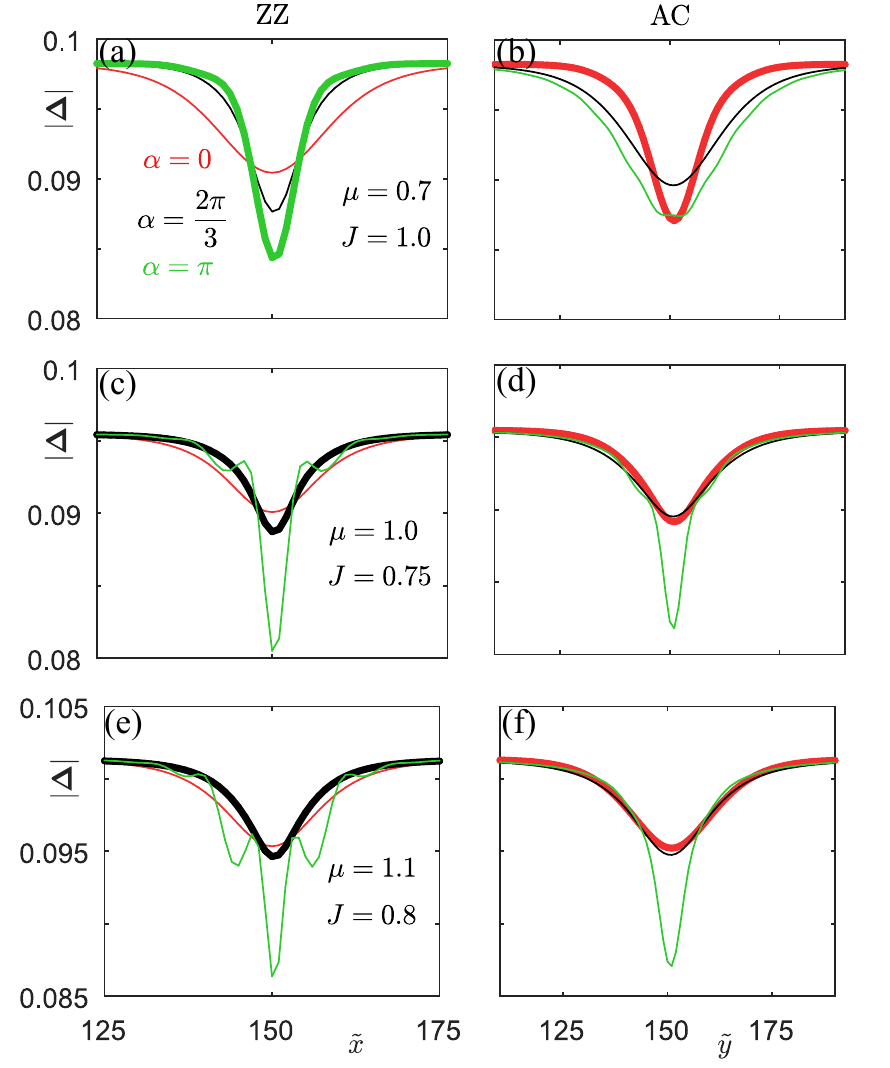}
	\caption[Strength]{(Color online.) Spatial profiles of the absolute values of the superconducting pairing $|\boldsymbol{\Delta|}$ for $\alpha = 0,2\pi/3, \pi$  along ZZ (a,c,e) and AC (b,d,f) direction with a domain wall at $\tilde{x} = \tilde{y} \approx 150$ for the same $\mu$ and $J$ as in Fig.~\ref{fig:FvsAlpha}(b,d,f). Thick lines indicate the configuration with lowest domain wall energy density.}
	\label{fig:Strength}
\end{figure}
To further understand the most favorable domain wall configurations we compare the profiles of the absolute values of the superconducting order parameter $|\boldsymbol{\Delta}|$ across AC and ZZ domain walls with $\alpha = 0, 2\pi/3, \pi$ (i.e. DWI, DWII at high-symmetry point, DWIII). These give the effective width of the domain walls in each configuration.
There is a clear suppression of $|\boldsymbol{\Delta}|$ at the domain wall in all cases, which implies that domain walls always carry an energy cost. Still, as we have shown above, these self-consistent profiles have lower energies than their corresponding sharp profiles which have no dip in $|\boldsymbol{\Delta}|$. Based on this, we directly understand why the exceptionally sharp and narrow domain wall order parameter profiles for $\alpha = \pi$ (green) in Figs.~\ref{fig:Strength}(c-f) give such an unfavorably high domain wall energy.
If we exclude these exceptionally narrow $\alpha = \pi$ configurations, we see in Fig.~\ref{fig:Strength} that, very generally, domain walls with the lowest energy (thick lines) also have the most narrow domain walls. Thus we can find the most favorable domain wall configuration by simply minimizing the domain wall width (apart from the situation where the domain wall gets disproportionally narrow).
One telling example is the ZZ domain wall with doping at the VHS, as displayed in Fig.~\ref{fig:Strength}(c). Here DWIII (green) is a very non-stable configuration (see Fig.~\ref{fig:FvsAlpha}(d)), which is also evident from the exceptionally narrow and even oscillatory order parameter profile, causing a notable increase in energy. 
When comparing DWI (red) and DWII (black) we see that DWI has a wider domain wall, with a slower recovery of superconductivity. This is a stable configuration, but it has a higher energy than the narrower DWII domain wall. 
Overall these results establish that domain wall configurations with the shortest healing length are energetically favorable. This can be seen as a consequence of robustness the chiral $d$-wave state.

%-------------------------------------------%
% Domain wall energy density minimum as function of mu
%-------------------------------------------%
\begin{figure}[htb]	
	\centering	
	\includegraphics[]{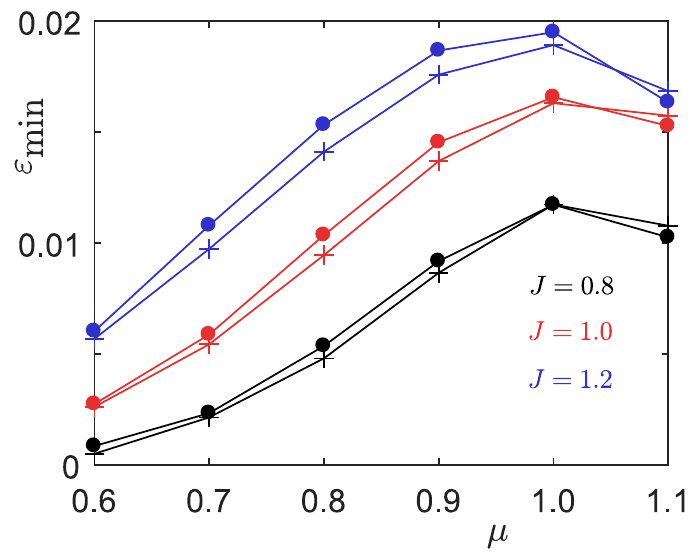}
	\caption[Character]{(Color online.) Domain wall energy density minimum $\varepsilon_\textrm{min}$ along ZZ (filled circle) and AC (cross) as a function of $\mu$ for $J = 0.8, 1.0, 1.2$.}
	\label{fig:FvsMU}
\end{figure}
Finally we report for both the AC and ZZ directions the minimum domain wall energy density $\varepsilon_\textrm{min}$ after fully relaxing the choice of $\alpha$ during the self-consistent procedure. 
We find that the competition between the AC and ZZ directions does not depend on $J$, as clearly seen Fig.~\ref{fig:FvsMU}. In terms of doping dependence, we find AC domain walls being favored at low $\mu$, while beyond the VHS ZZ domain walls are preferred. The fact that the system does not have a preferred direction at the VHS can be understood from the perfect hexagonal structure of the Fermi surface at that point. The VHS also sees the largest domain wall energies, which is due to the chiral $d$-wave state being strongest at that point \cite{nandkishore2012chiral, Kiesel12, Wang11}. 

%------------------------------------%
% DW WITH STABLE ENERGY
%------------------------------------%
\subsection{Properties of minimum energy domain walls} 
\begin{figure}[ht!]	
	\centering	
	\includegraphics[scale =0.5]{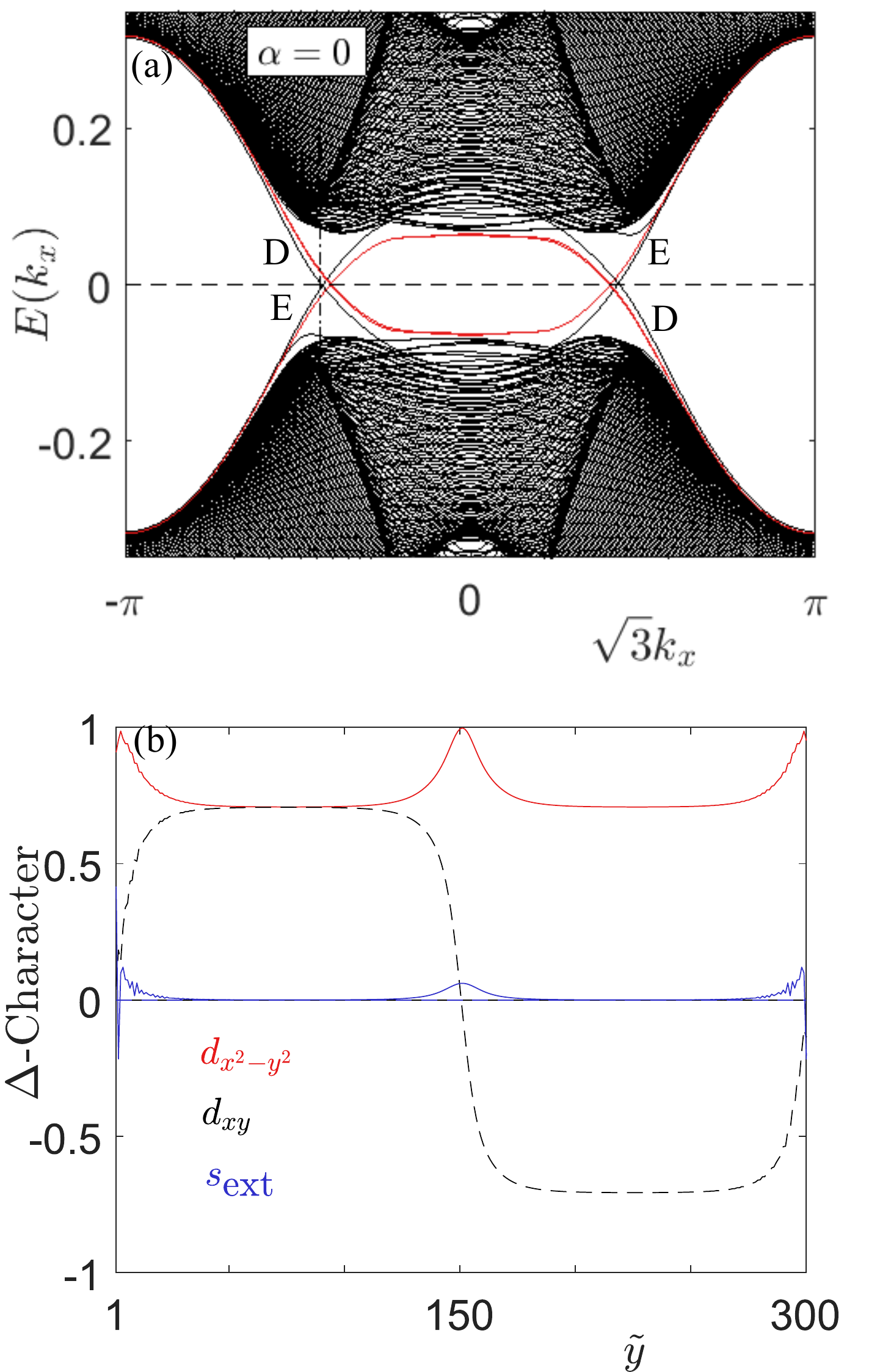}
	\caption[Spectrum]{(Color online.) Spectrum (a) and spatial profile of the imaginary (dashed) and real (solid) parts of the $\Delta$-character (b) for the AC DWI; the most stable domain wall at low doping levels. Low energy states within the gap in (a) are for self-consistent (black) and sharp domain walls (red), while the dashed horizontal line indicates the Fermi level.  E and D denote edge and domain wall states, respectively, which are both doubly degenerate. Vertical line shows the $k$-point for which Fig.~\ref{fig:Prob} is obtained. Here $\alpha=0$, $\mu =0.7$, $J=1.0$.}
	\label{fig:SpectrumAC}
\end{figure}

\begin{figure}[ht!]	
	\centering	
	\includegraphics[scale=0.5]{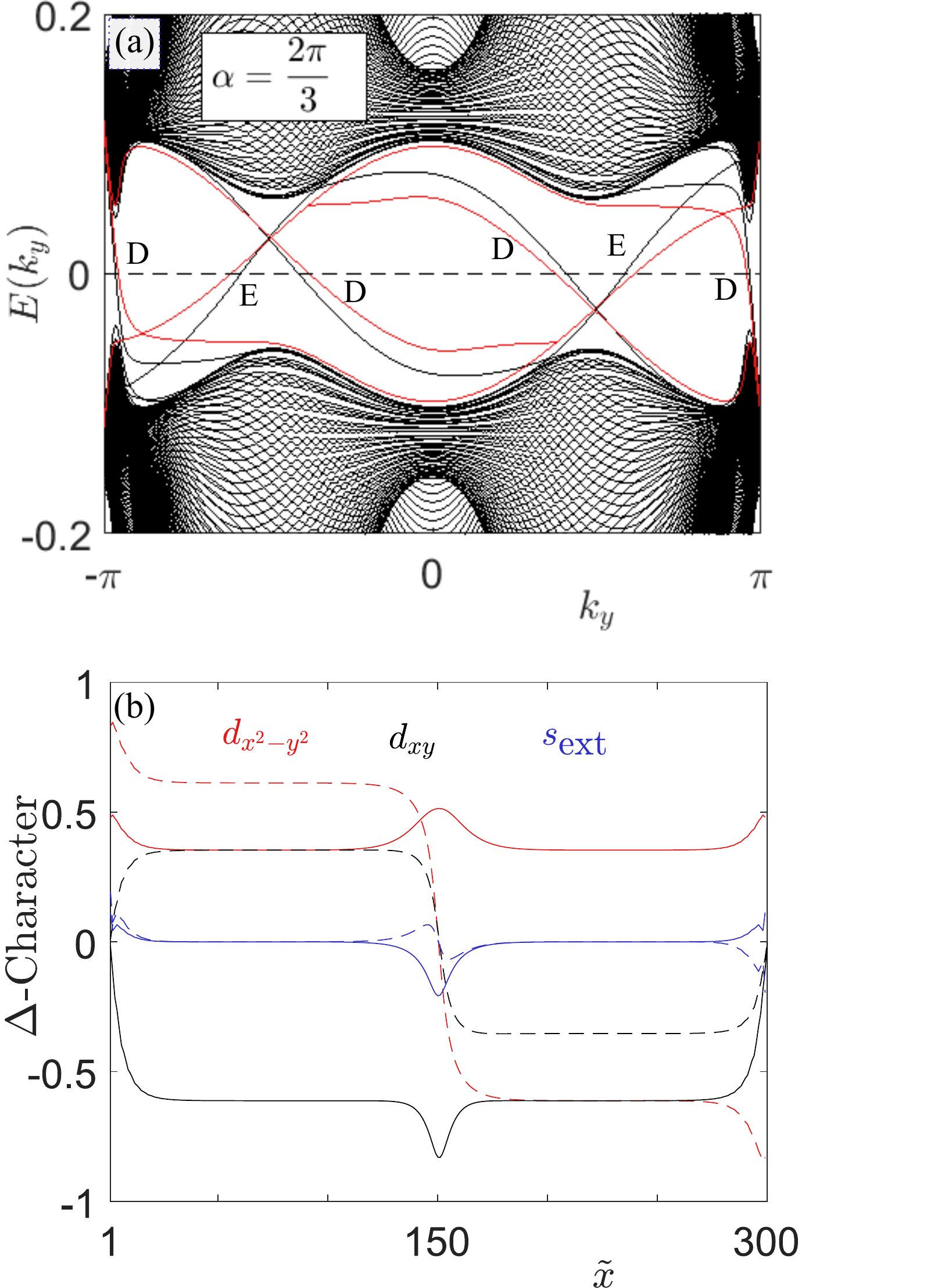}
	\caption[Spectrum]{(Color online.) Spectrum (a) and spatial profile of the imaginary (dashed) and real (solid) parts of the $\Delta$-character (b) for the ZZ DWII; the most stable domain wall at doping beyond the VHS. Low energy states within the gap in (a) are for self-consistent (black) and sharp domain walls (red), while the dashed horizontal line indicates the Fermi level. E and D denote doubly degenerate edge and singly degenerate domain wall states, respectively. Here $\alpha=\frac{2\pi}{3}$, $\mu =1.1$, $J=0.8$.}
	\label{fig:SpectrumZZ}
	\end{figure}

Having established the most stable configuration for chiral $d$-wave domain walls as a function of doping, we here focus on the properties of these domain walls. We study both the quasiparticle excitation spectrum with its protected subgap domain wall states and the character of the order parameter at and around the domain wall. 
We display the spectrum and the $\Delta$-character for the AC DWI at low $\mu$ in Fig.~\ref{fig:SpectrumAC} and for the ZZ DWII beyond VHS in Fig.~\ref{fig:SpectrumZZ}.
%
%Figure of spectrum and order parameter character for DW with minimum energy along AC
%
%Figure of spectrum and order parameter character for DW with minimum energy along ZZ
In the vicinity of the VHS the spectrum and $\Delta$-characters for both directions are qualitatively the same as the figures reported here, and thus these two situations cover all different behaviors. Moreover, we find that the overall form of the $\Delta$-character is only affected by $\alpha$ and does not depend on other parameters or the domain wall direction.

%-------------------------------------------------------------%
% Edge states
%-------------------------------------------------------------%
The non-trivial topology with $C = \pm 2$ in the bulk of a chiral $d$-wave superconductor always gives rise to two protected and chiral, or co-propagating, edge states at any outer boundary of the superconductor. These states are marked with E in the spectrum plots, Figs.~\ref{fig:SpectrumAC}(a),\ref{fig:SpectrumZZ}(a). Both domains in our setup have edge states with positive dispersion. This is because in the left domain the chirality gives clockwise rotating edge states, while in the right domain the edge states rotates counter-clockwise. Since the outer edges are very well separated spatially and their termination is symmetric, these states appear as perfectly doubly degenerate in the spectrum.
At each outer edge we also see how the character of the order parameter becomes purely $d_{x^2-y^2}$-wave for both the AC and ZZ domain walls, see Figs.~\ref{fig:SpectrumAC}(b),\ref{fig:SpectrumZZ}(b). This is fully consistent with earlier results studying outer edges,\cite{Black-Schaffer12PRL} as well as crystallographic rearrangements causing effective edge physics.\cite{schmidt2016chiral}

\begin{figure}[htb]	
	\centering	
	\includegraphics[scale =0.8]{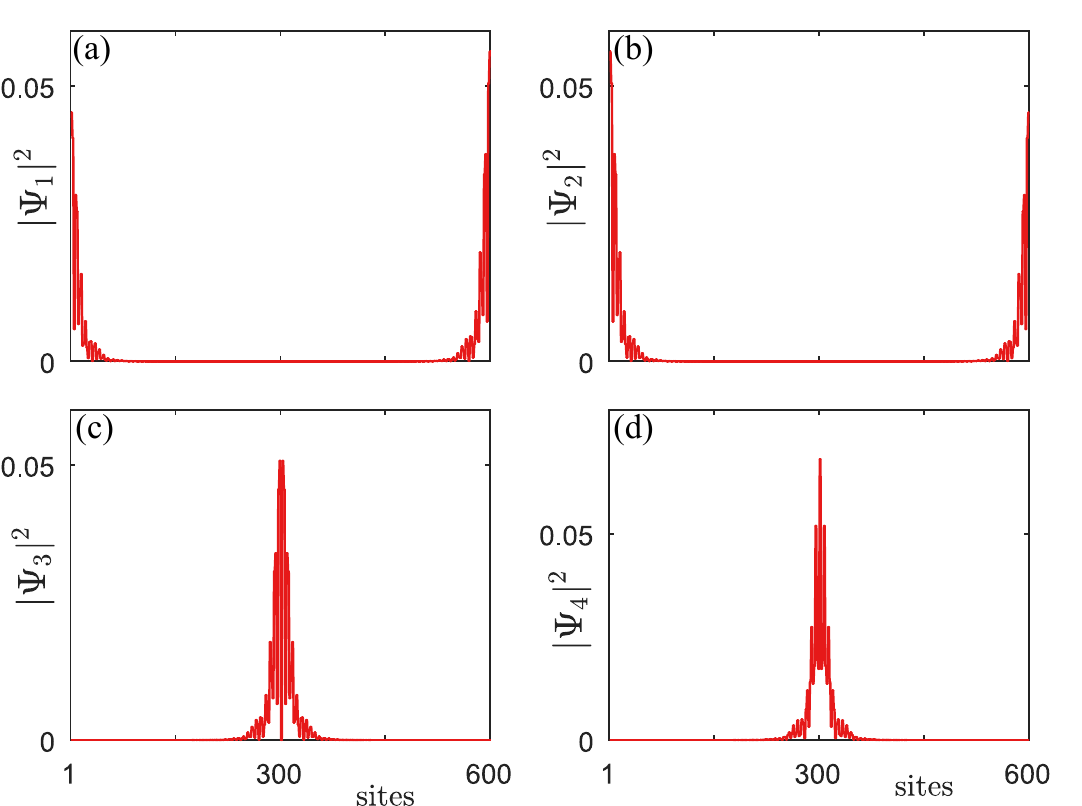}
	\caption[Spectrum]{(Color online.) Probability density of the four states closest to zero energy as a function of distance from the domain wall with E edge states (a,b) and D domain wall states (c,d) for the AC DWI at the $k_x$ value indicated by a vertical line in Fig.~\ref{fig:SpectrumAC}(a).}
	\label{fig:Prob}
\end{figure}

The remaining subgap states are associated with the domain wall and marked with a D, as also confirmed in Fig.~\ref{fig:Prob}. At the domain wall the winding number changes by 4 which, according to the bulk-edge correspondence,\cite{Volovik97, HassanKane2010RMP, qi2011topological, graf2013bulk} gives rise to a total of four localized domain wall states. Fully consistent with how the edge states rotate in each domain. These domain wall states all have a negative dispersion, as seen in Figs.~\ref{fig:SpectrumAC}(a),\ref{fig:SpectrumZZ}(a). 
Since all four domain wall states overlap in real space, they can easily avoid energy degeneracies. This is clearly the case for the ZZ domain wall, where there are four well separated D states. However, for the AC domain wall there is still an energy degeneracy between the D states. This degeneracy can be understood by noting that the order parameter at the edge and at the domain wall is virtually identical for the AC DWI domain wall: both show an almost pure $d_{x^2-y^2}$ symmetry, with the $d_{xy}$-wave being extinguished due to the pair breaking of any honeycomb edge.\cite{Black-Schaffer12PRL} Thus the D and E states for the AC domain wall system should be very similar (but with opposite dispersion), as we also find in Fig.~\ref{fig:SpectrumAC}(a). This is in sharp contrast to the situation at finite $\alpha$, as is always the case for ZZ domain walls. Here the domain wall has an order parameter with an unavoidable mixture of the $d_{x^2-y^2}$ and $d_{xy}$ symmetries. Thus the D states should behave differently from the E state, since at the edge there is almost pure $d_{x^2-y^2}$ symmetry, independent on $\alpha$. This difference between the D and E states naturally grows as  $\alpha$ increases, resulting in the substantial spectral disparity between the E and D states in the ZZ DWII system in Fig.~\ref{fig:SpectrumZZ}(a).
Interestingly, the ZZ DWII subgap domain wall states have a very steep dispersion close to $k_y =\pm \pi$. 
This means these states contribute vanishingly small energy to the total domain wall energy density $\varepsilon$, which enables the stability of this configuration. This offers an explanation to why the ZZ domain walls usually have an energy preference for configurations with $\alpha$ not at a high symmetry value (i.e. different for $0,2\pi/3,\pi$). It is simply the intricate details of the subgap domain wall spectrum that drives the energetics of the domain wall.
Apart from the dominating $d$-wave terms, there is also a small $s_\textrm{ext}$-wave component at both edges and domain walls in all cases we have studied.

%-------------------------------------------------------------%
% Probability density
%-------------------------------------------------------------%
In Fig.~\ref{fig:Prob} we confirm the spatial properties of the four states closest to zero in Fig.~\ref{fig:SpectrumAC}(a) at the $k_x$-point marked with a vertical line. In Fig.~\ref{fig:Prob}(a,b) we plot the wave function probability density for the two energy degenerate states, here called $\Psi_{1,2}$,  marked with E in Fig.~\ref{fig:SpectrumAC}(a). Clearly these two states reside at the edges of the system. By a simple rotation in eigenvector space it is always possible to disentangle these two eigenstates into their properly spatially separated components.
Similarly, the two D marked states, $\Psi_{3,4}$, are plotted in Fig.~\ref{fig:Prob}(c,d). They are linear combinations of two of the total four domain wall states (the other two domain wall state exists at $-k_x$) and clearly resides at the domain wall and in both domains. Due to their significant spatial overlap it is usually not possible to isolate one from the other when the states are energy degenerate. 
%Figure of wavefunction
 
%------------------------------------%
%DOMAIN WALL WIDTH
%------------------------------------%
\subsection{Recovery length}
\begin{figure}[htb]	
	\centering	
	\includegraphics[]{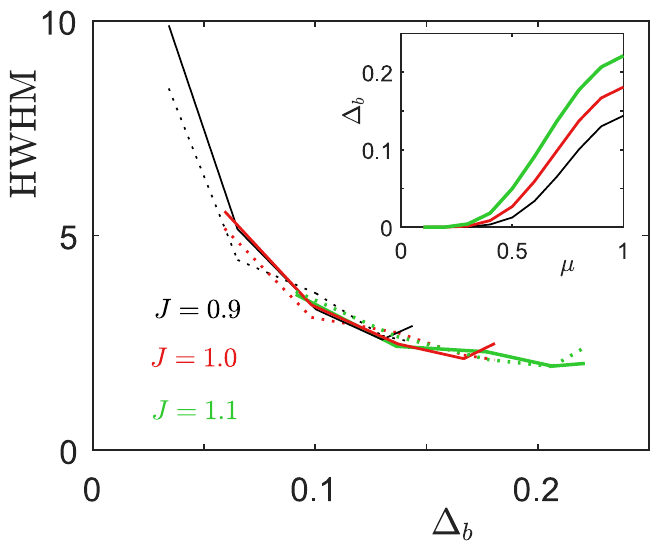} 
	\caption[FWHM]{(Color online.) Half width at half maximum (HWHM) of domain wall at $\alpha_\textrm{min}$ for ZZ (dotted) and AC (solid) domain walls as a function of $\Delta_b$ for several values of $J$. Inset shows the relationship of $\Delta_b$ with $\mu$.}
		\label{fig:WidthMinZZAC}
\end{figure}

Finally, we investigate how superconductivity recovers from its suppression at the domain wall. For this purpose we extract the half width at half maximum (HWHM) of the suppression of the order parameter at the domain wall. We plot this quantity for the AC and ZZ domain walls at $\varepsilon_\textrm{min}$ for several coupling strengths and doping levels in Fig.~\ref{fig:WidthMinZZAC}. The errors from the fitting used to extract HWHM are negligible and ignored in the plot.
 We find that superconductivity recovers rather quickly away from the domain wall. This short recovery, or healing, length signifies that chiral $d$-wave state is very robust and the result is in agreement with earlier results from defects\cite{Lothman14} and edges\cite{Black-Schaffer12PRL} in heavily doped superconducting graphene. 
 Surprisingly, the behavior of the domain wall width as a function of the strength of the bulk superconducting state, $\Delta_b$, is universal, irrespective of system parameters or even domain wall direction. The small discrepancy at very weak $\Delta_b$ are diminished by increasing the system size to accommodate the longer superconducting coherence length.  The small increase in HWHM at high $\mu$ is due to the increased competition from the $s_\textrm{ext}$-wave state. 
Thus knowing the strength of the order parameter in the bulk is enough to determine the width of the domain wall in a chiral $d$-wave superconductor, independent on domain wall direction in the lattice.
 The behavior of HWHM with the bulk strength of superconductivity is an important result in that it recovers the generic relationship between the recovery length and order parameter, $\xi \sim \Delta^{-1}$, that is obtained from phenomenological theory. 
 
%------------------------------------%
%CONCLUSION
%------------------------------------%
\section{Concluding discussion}
%------------------------------------%
% Summary
%------------------------------------%
In this work we have performed a self-consistent study of domain walls in the chiral $d$-wave superconducting state on the honeycomb lattice, studying general domain wall directions as well as global phase shifts across the domain wall.
At doping levels up to the VHS, AC domain walls with a global zero phase are energetically most favorable, while at doping levels beyond the VHS, ZZ domain walls are most stable and then with an additional global phase of $\alpha \sim 2\pi/3$. The most stable domain wall configurations are generally the solutions with the narrowest suppression of the order parameter, i.e.~smallest domain wall width, although exceptionally deep and narrow domain wall profiles have notably higher energy. In fact, sharp domain wall order parameter profiles used in non-self-consistent calculation almost always predicts the wrong lowest energy domain wall configuration.
The non-universality of $\alpha_{\rm min}$ that we find has also been found in chiral $p$-wave superconductors.\cite{sigrist1999role}
Accompanying the domain wall formation are four domain wall states, satisfying the bulk-boundary correspondence principle\cite{Volovik97,HassanKane2010RMP, qi2011topological, graf2013bulk} and propagating in opposite direction to the system edge states. The domain wall state spectrum are notably different from the edge state spectrum for $\alpha$ non-zero. This is due to the mixing of the $d$-wave characters at the domain wall, which does not occur at any of the outer edges, and also helps stabilize domain walls for $\alpha$ not at a high symmetry value.
Moreover, we have established that, regardless of the preferred domain wall direction, the recovery of bulk superconductivity from the domain wall is fast and universal. The recovery length which is only depending on the strength of the bulk superconductivity following an inverse relationship.

%------------------------------------%
% Connection to experiment
%------------------------------------%
Our results can be experimentally verified in tunneling experiments. The domain wall states result in non-vanishing local DOS, or differential conductance, in the bulk at zero energy in chiral $d$-wave superconductors. However, subgap states due to potential defects and even vacancies in a chiral $d$-wave superconductor have been shown to never reach zero energy.\cite{Lothman14} This implies that zero energy states away from sample edges directly indicate the presence of domain wall states and thus a scanning probe can be used to track the domain wall structure.
We do not expect disorder to strongly affect these results. Even moderately high (Anderson) disorder has  been shown to not significantly influence the chiral $d$-wave edge states.\cite{Black-Schaffer12PRL} Since the domain wall states and edge states emerge from the same topological phenomenon, we do not anticipate disorder to notably influence the domain wall states. 

% -------------------------------------------------- %
% ACKNOWLEDGMENTS
% -------------------------------------------------- %
\acknowledgments
We are grateful to J.~Cayao, D.~Kuzmanovski, T.~L\"othman, and M.~Sigrist for discussions and thank the Swedish research council (VR), the Swedish Foundation for Strategic Research (SSF), the G\"oran Gustafsson Foundation, and the Knut and Alice Wallenberg Foundation through the Wallenberg Academy Fellows program for financial support.

\bibliography{abbrRef,DWref}

\end{document}